\documentclass[a4paper, superscriptaddress, amsfonts, amssymb, amsmath, reprint, showkeys, nofootinbib, twoside,aps,prl]{revtex4-1}

\usepackage{amsmath}

\usepackage{xcolor}
\usepackage{graphicx}
\usepackage{dcolumn}
\usepackage{bm}
\usepackage{amsmath}
\usepackage{mathtools}
\usepackage{enumitem}
\usepackage{xcolor}
\usepackage{graphicx}
\usepackage{dcolumn}
\usepackage{bm}
\usepackage{float}
\usepackage{lipsum}

\begin{document}

\preprint{APS/123-QED}

\title{Predictive Information Decomposition as a Tool to Quantify Emergent Dynamical Behaviors In Physiological Networks}

\author{Luca Faes}
\affiliation{Department of Engineering, University of Palermo, Italy}
\email{luca.faes@unipa.it}
 \affiliation{Faculty of Technical Sciences, University of Novi Sad, Serbia}

\author{Gorana Mijatovic}
 \affiliation{Faculty of Technical Sciences, University of Novi Sad, Serbia}
 \email{gorana86@uns.ac.rs}

 \author{Laura Sparacino}
\affiliation{Department of Engineering, University of Palermo, Italy}
\email{laura.sparacino@unipa.it}

\author{Alberto Porta}
 \affiliation{Department of Biomedical Sciences for Health, University of Milano, Italy}
  \affiliation{Department of Cardiothoracic, Vascular Anesthesia and Intensive Care, IRCCS Policlinico San Donato, Italy}


\begin{abstract}
Objective: This work introduces a framework for multivariate time series analysis aimed at detecting and quantifying collective emerging behaviors in the dynamics of physiological networks. 
Methods: Given a network system mapped by a vector random process, we compute the predictive information (PI) between the present and past network states and dissect it into amounts quantifying the unique, redundant and synergistic information shared by the present of the network and the past of each unit. Emergence is then quantified as the prevalence of the synergistic over the redundant contribution. The framework is implemented in practice using vector autoregressive (VAR) models.
Results: Validation in simulated VAR processes documents that emerging behaviors arise in networks where multiple causal interactions coexist with internal dynamics.
The application to cardiovascular and respiratory networks mapping the beat-to-beat variability of heart rate, arterial pressure and respiration measured at rest and during postural stress reveals the presence of statistically significant net synergy, as well as its modulation with sympathetic nervous system activation.
Conclusion: Causal emergence can be efficiently assessed decomposing the PI of network systems via VAR models applied to multivariate time series. This approach evidences the synergy/redundancy balance as a hallmark of integrated short-term autonomic control in cardiovascular and respiratory networks.
Significance: Measures of causal emergence provide a practical tool to quantify the mechanisms of causal influence that determine the dynamic state of cardiovascular and neural network systems across distinct physiopathological conditions.
\end{abstract}

\maketitle

\section{Introduction}
\label{sec:introduction}
In recent years, the development of biomedical signal processing methods has seen a shift towards the multivariate analysis of physiological time series collected simultaneously from the biological system of interest. This approach is indispensable in the fields of network neuroscience and physiology \cite{bassett2017network,ivanov2021new}, where physiological systems (e.g., the brain or the cardiovascular and respiratory systems) are represented as networks formed by highly interconnected units (e.g., different brain areas or organ systems) which exhibit complex dynamics at multiple spatial and temporal scales. In these fields, multivariate time series analysis methods are frequently applied, for instance in the brain to study functional connectivity across different modalities \cite{he2019electrophysiological}, or in cardiovascular physiology to assess the homeostatic physiological mechanisms reflected by the variability of heart period (HP), arterial pressure (AP) and respiration (RESP) \cite{schulz2013cardiovascular}.

In the multivariate analysis of physiological time series, an important issue is how to detect and dissect the variety of complex collective behaviors that can emerge out of the interaction among the multiple units composing the observed network. In this context, two main approaches are used for the analysis of dynamic network systems.
The first consists in characterizing \textit{connectivity} by computing a big variety measures of the interaction between pairs of time series (see, e.g., \cite{porta2015wiener,cliff2023unifying}) and then analyzing the graph of the resulting pairwise connections with properly defined “network measures” \cite{sporns2018graph}. The second approach investigates \textit{integration} through the computation of one-dimensional (single-value) metrics (see, e.g., \cite{barrett2011practical,oizumi2016unified}) obtained comparing the collective network dynamics with the dynamics of individual (groups of) units. In spite of their widespread diffusion, these two approaches present some shortcomings that limit their usability for the collective analysis of network behaviors: pairwise connectivity methods neglect high-order interactions possibly shaping the overall network dynamics \cite{majhi2022dynamics}, while integration measures are difficult to interpret in empirical network analyses because they have been shown to behave inconsistently \cite{mediano2018measuring}. 

Here, we argue that the two main approaches to multivariate time series analysis described above are inherently limited as regards the assessment of collective emergent behaviors in network systems: connectivity analysis quantifies the information shared between pairs of units, thus putting focus on the links rather than on the whole system;
the analysis of integration quantifies the excess of information generated by the whole system over that generated by its parts, thus putting focus on the concept of integration rather than to that of emergence.
On the other hand, according to the recent theory of causal emergence \cite{rosas2020reconciling}, measures of emergent behaviors should investigate whether and how the future state of a network system can be predicted better from the collective dynamics of all units, rather than from the individual dynamics of single units, as a result of synergistic high-order interactions. Following this intuition, the present work introduces an information-theoretic method to assess, in dynamic network systems mapped by vector random processes, how collective behaviors emerge in the transition across network states.
To do this, we leverage information decomposition strategies \cite{timme2014synergy,williams2010nonnegative} applied to the random variables that map the temporal evolution of the analyzed system.
Specifically, we decompose the predictive information \cite{bialek2001predictability} shared between the present state of the whole system (vector ‘target’ variable) and the past states of each subsystem (‘source’ variables)  in order to assess the balance between synergistic and redundant high-order interactions.
In the frame of information theory, redundancy and synergy measure respectively the information conveyed to the target simultaneously by all sources, and the information conveyed by all sources together but none of them individually. Here, the synergy/redundancy balance is computed following two alternative approaches: the whole-minus-sum (WMS) decomposition \cite{timme2014synergy}, which provides a direct calculation by elaborating mutual information terms; and the partial information decomposition (PID) \cite{williams2010nonnegative} which quantifies separately the synergistic and the redundant information terms together with the unique information that is held by each source but not by the others.

The proposed approach is formalized in the so-called framework of Predictive Information Decomposition (PrID), 
which is implemented in a data-efficient way in the context of vector autoregressive (VAR) modeling of multivariate time series \cite{faes2016information,mijatovic2024assessing}, and tested in theoretical examples to illustrate how it allows to relate emerging collective behaviors with the causal mechanisms shaping the connectivity structure of the analyzed network. Then, it is employed in a paradigmatic application in the field of computational physiology, i.e. the study of cardiovascular and respiratory networks probed measuring the time series of HP, systolic and diastolic AP, and RESP variability in different physiological conditions \cite{javorka2017basic}. While these networks have been largely studied using pairwise and higher-order measures \cite{faes2016information,javorka2017basic,porta2021categorizing,mijatovic2024assessing}, here they are analyzed for the first time from the perspective of causal emergence.


\section{Predictive Information Decomposition}
\label{sec:PrID}
Given a network system $\mathcal{X}$ composed by $N$ interacting units, with dynamical activity described by the stationary vector random process $X=\{X^1,\ldots,X^N \}$, let us denote as $X_n=[X^1_n \cdots X^N_n]^\intercal$ and $X_{<n}=[X^1_{<n} \cdots X^N_{<n}]^\intercal$ the vector random variables sampling the processes at the present time $n$ and at the past times $n-1,n-2,\ldots$, respectively (see Fig. 1a for a graphical representation). In the framework of information dynamics, the total amount of Shannon information that is transferred through the analyzed processes from past to present is given by the mutual information (MI) between the present and past states, denoted as $I(X_n;X_{<n})$ \cite{bialek2001predictability}. This measure, which we denote as \textit{predictive information} (PI), quantifies the degree of predictability of the present state of the network given its own past states; it extends to multivariate processes the information storage of the $i^{\mathrm{th}}$ scalar process, $I(X^i_n;X^i_{<n})$ \cite{lizier2012local}.
In this work, to assess how the current state of the analyzed system emerges from the joint temporal evolution of its subsystems, we perform PrID decomposing the PI into terms quantifying the contribution of the history of each individual process (described by the \textit{source} variable $X^i_{<n}$, $i=1,\ldots,N$) to the present state of the system (described by the \textit{target} variable $X_n$), and terms quantifying how the $N$ source variables in $X_{<n}$ interact collectively to determine the target variable $X_n$ through high-order effects. If such high-order effects have synergistic nature, the behavior of the network system is regarded as emergent \cite{rosas2020reconciling}.

\begin{figure*} [t!]
    \centering
    \includegraphics[scale= 1.05]{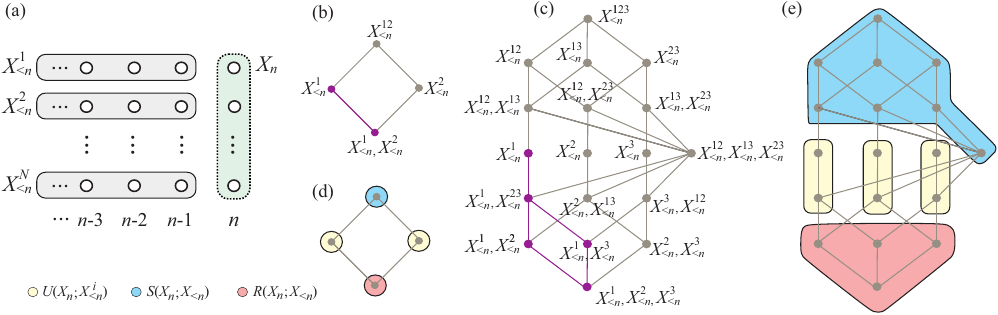}
    \caption{Graphical depiction of PID-based Predictive Information Decomposition. (a) Schematic of the random variables collected to describe the present state of the analyzed network system (green) and the past states of each constituent unit (gray). (b,c) Standard PID lattices defined for $N=2$ and $N=3$ sources, depicting the information atoms and the ordered connections identifying the atoms to be summed to determine the redundancy function (purple: atoms determining the redundancy of the first source, $I_{\cap}(\{1\})=I(X_n;X^1_{<n})$). (d,e) Coarse graining of the PID atoms identifying the unique (yellow), synergistic (blue) and redundant (red) components providing an exact decomposition of PI.}
    \label{intro_PrID}
\end{figure*}

\subsection{Whole-minus-sum decomposition}
To decompose the PI, we first consider a so-called \textit{whole-minus-sum} (WMS) approach, developed following the principle of many information-theoretic approaches to multivariate analysis which compare the values of an information measure when it is computed for the whole system or is summed over its parts \cite{timme2014synergy}. This strategy leads us to decompose the PI as 
\begin{equation}
    I(X_n;X_{<n})=\sum_{i=1}^{N}I(X_n;X^{i}_{<n}) +  \Delta_{\mathrm{WMS}}(X_n;X_{<n}),
   \label{WMSdec}
\end{equation}
where $I(X_n;X^{i}_{<n})$ quantifies the information shared between the present state of the whole network and the past states of its $i^{\mathrm{th}}$ unit. The term $\Delta_{\mathrm{WMS}}(X_n;X_{<n})$ arises from the fact that the PI is generally different than sum of the MI terms. This quantity reflects the \textit{synergy/redundancy} balance of the informational circuits linking the target $X_n$ to the sources $X^i_{<n}$: if the past states of $X$ contribute more to its present when they are taken separately we have $\Delta_{\mathrm{WMS}}<0$, denoting prevalence of redundancy; if on the contrary the present of $X$ is better described considering its own past as a whole we have $\Delta_{\mathrm{WMS}}>0$, denoting prevalence of synergy and thus presence of emergent behaviors in the network.

The decomposition (\ref{WMSdec}) has the advantage of simplicity, as it can be implemented solely by computing MI terms. Its drawback is that it conflates synergy and redundancy in one single measures and, as we will see in the next sections, it overestimates redundancy as the network size grows.

\subsection{Partial information decomposition: formulation}
As an alternative to (\ref{WMSdec}), we propose to decompose the PI through a PID approach \cite{williams2010nonnegative}:
\begin{equation}
    I(X_n;X_{<n})=\sum_{i=1}^{N}U(X_n;X^{i}_{<n})+R(X_n;X_{<n})+S(X_n;X_{<n}),
   \label{PIDdec}
\end{equation}
where the terms $R(X_n;X_{<n})$ and $S(X_n;X_{<n})$ explicitly quantify the \textit{redundant} and \textit{synergistic} information brought to the target variable $X_n$ by the source variables collected in $X_{<n}$, while $U(X_n;X^{i}_{<n})$ quantifies the \textit{unique} information brought to $X_n$ by the $i^{\mathrm{th}}$ source $X^{i}_{<n}$.
Besides (\ref{PIDdec}), the PID formulates a set of $N$ additional equations stating that the information shared between each source and the target is in part unique to that source and in part redundant with the other sources:
\begin{equation}
    I(X_n;X^{i}_{<n})=U(X_n;X^{i}_{<n})+R(X_n;X_{<n});
   \label{MIdec}
\end{equation}
note that the $N+1$ equations in (\ref{PIDdec},\ref{MIdec}) do not suffice to determine unequivocally the $N+2$ unknown quantities evidenced by (\ref{PIDdec}). This indeterminacy is addressed by solving the PID as shown in the next subsection.

Since the PID (\ref{PIDdec}) provides separate redundant and synergistic contributions, it allows to straightforwardly quantify the balance between synergy and redundancy as
\begin{equation}
    \Delta_{\mathrm{PID}}(X_n;X_{<n})=S(X_n;X_{<n})-R(X_n;X_{<n}).
   \label{PIDdelta}
\end{equation}
While the presence of strictly positive synergy is a sufficient condition for a network system to exhibit emergent behavior \cite{rosas2020reconciling}, here we take the balance $\Delta_{\mathrm{PID}}$ as a "strong" measure of  emergence. 
Crucially, even if $\Delta_{\mathrm{PID}}$ is conceptually similar to the measure $\Delta_{\mathrm{WMS}}$ appearing in (\ref{WMSdec}), it gives a better account of the balance between synergistic and redundant higher-order interactions in the analyzed system. This is demonstrated  inserting (\ref{PIDdec}) and (\ref{MIdec}) in (\ref{WMSdec}) to derive
\begin{equation}
    \Delta_{\mathrm{WMS}}(X_n;X_{<n})=S(X_n;X_{<n})-(N-1)R(X_n;X_{<n}),
   \label{double_counting}
\end{equation}
which shows that $\Delta_{\mathrm{WMS}}$ is equivalent to $\Delta_{\mathrm{PID}}$ only when $N=2$ sources are considered, while it counts redundancy multiple times for $N \geq 3$.

\subsection{Partial information decomposition: solution}
The PID decomposition (\ref{PIDdec}) is appealing because it puts in evidence non-negative terms which can be related to how $X_n$ emerges from the interactions among the components of $X_{<n}$. However, contrary to the WMS decomposition (\ref{WMSdec}), the information terms of the PID identified on the r.h.s. of (\ref{PIDdec}) cannot be obtained using only classical Shannon's information measures. 
To solve the PID formulated in (\ref{PIDdec},\ref{MIdec}), we follow the seminal works of Williams and Beer \cite{williams2010nonnegative} and Rosas et al. \cite{rosas2020reconciling}. Specifically, the PID is addressed by constructing a so-called redundancy lattice over the set of all combinations of source variables \cite{williams2010nonnegative}, and then aggregating the information associated to specific elements of such set according to a so-called "coarse graining" approach \cite{rosas2020reconciling}.

The PID decomposes the MI between the target variable and the set of $N$ source variables as the sum of non-negative terms (information atoms); each atom identifies a combination of sources taken separately or together. In our case, being $X_n$ the target and $X^i_{<n}, i=1,\ldots,N$, the sources, the PID realizes the following decomposition of the PI:
\begin{equation}
    I(X_n;X_{<n})=\sum_{\alpha \in \mathcal{A}}I_{\delta}(X^{\alpha}_{<n}),
   \label{PIDatoms}
\end{equation}
where $I_{\delta}(X^{\alpha}_{<n})$ is the information brought to $X_n$ by the atom $\alpha$, and the set of all atoms $\mathcal{A}$ collects all the subsets of $\{1,\ldots N\}$ where no elements is contained in another \cite{williams2010nonnegative}.
Figs. 1b and 1c show the lattice structure for the cases of $N=2$ and $N=3$ processes. 

Besides the lattice, the PID defines a so-called redundancy function which quantifies the amount of overlapping information about the target that is common to a set of sources. For a given atom $\alpha \in \mathcal{A}$, the redundancy is the sum of the information associated to $\alpha$ and all the atoms connected to it downwards in the lattice:
\begin{equation}
    I_{\cap}(X^{\alpha}_{<n})=\sum_{\beta \preceq \alpha}I_{\delta}(X^{\beta}_{<n});
   \label{RedAtoms}
\end{equation}
examples of computation of $I_{\cap}(X^1_{<n})$ for the cases $N=2$ and $N=3$ are reported in Fig. 1b and Fig. 1c, respectively.
Given (\ref{RedAtoms}), the information associated with the atom $\alpha$ can be derived straightforwardly from the redundancy of that atom and from the information of the atoms positioned below in the lattice, i.e. $I_{\delta}(X^{\alpha}_{<n})=I_{\cap}(X^{\alpha}_{<n})-\sum_{\beta \prec \alpha}I_{\delta}(X^{\beta}_{<n})$, thus providing an iterative way to derive the information of the various atoms. In practice, the information associated with all atoms is obtained through a Moebius inversion applied to the redundancy values \cite{williams2010nonnegative}.

Thus, the PID is solved once a measure of redundant information $I_{\cap}(\cdot)$ is specified. Among several possibilities, in this work where we implement our framework focusing on linear Gaussian systems (see Sect. \ref{sec:Computation}) we follow the minimum MI (MMI) PID strategy \cite{barrett2015exploration}. Accordingly, we define the redundant information of the generic atom containing $J$ subsets of sources, $\alpha=\{\alpha_1,\ldots \alpha_J \} \in \mathcal{A}$, as the minimum of the MI shared by the past of the subsystems identified by each subset $\alpha_j$, i.e. $X^{\alpha_j}_{<n}$, and the present of the whole system:
\begin{equation}
   I_{\cap}(X^{\alpha}_{<n})=\min_{j=1,\ldots,J}I(X_n;X^{\alpha_j}_{<n}).
   \label{Red}
\end{equation}

Once the information of each atom is computed, it needs to be associated to the unique, redundant and synergistic terms in (\ref{PIDdec}) to complete the decomposition. To do so, we coarse-grain the PID summing $I_{\delta}(X^{\alpha}_{<n})$ over the atoms satisfying the criteria defined in \cite{rosas2020reconciling} (first order coarse-graining): $U(X_n;X^{i}_{<n})$ is the information carried by $X^{i}_{<n}$ about $X_n$ that no other source $X^{j}_{<n}$, $j \neq i$, has on its own; $R(X_n;X_{<n})$ is the information that is held by at least two sources $X^{i}_{<n}$ and $X^{j}_{<n}$, $i \neq j$, taken individually; $S(X_n;X_{<n})$ is the information that is not carried by any individual source $X^{i}_{<n}$ when considered separately from the rest. These aggregation rules result in the unique, redundant and synergistic information shown in Fig. 1d,e; note that, while for the case of two sources there is one-to-one correspondence between the terms in (\ref{PIDdec}) and in (\ref{PIDatoms}), the coarse graining is crucial when more sources are considered: for $N=3$, each unique term sums the information of two atoms and the synergistic and information terms gather information respectively from 8 and 4 atoms.

\section{Practical Computation}
\subsection{Formulation for Gaussian Systems}
\label{sec:Computation}
In this work, the practical implementation of the PI decompositions defined in Sect \ref{sec:introduction} is performed exploiting the exact formulation for the MI holding for multivariate processes with Gaussian distribution \cite{faes2016information,mijatovic2024network,mijatovic2024assessing}.
Specifically, if the variables sampling the zero-mean vector process $X$ have a joint Gaussian distribution, the statistical dependencies between them are fully accounted by the vector autoregressive (VAR) model of order $p$ defined as
\begin{equation}
   X_n = \sum_{k=1}^{p}\mathbf{A}_k X_{n-k} + U_n
   \label{VAR},
\end{equation}
where $\mathbf{A}_k$ is a $N \times N$ matrix model coefficients relating the present of the processes with their past at lag $k$, and $U$ is an $N-$dimensional white noise process with covariance matrix $\Sigma_U$.
Given the representation in (\ref{VAR}), the PI can be formulated analytically as \cite{barrett2011practical,mijatovic2024network}
\begin{equation}
    I(X_n;X_{<n})=\frac{1}{2} \log \frac{|\Sigma_{X}|}{|\Sigma_{U}|},
   \label{PIgauss}
\end{equation}
where $\Sigma_{X}$ is the covariance matrix of the original process $X$.

Moreover, starting from the full VAR representation (\ref{VAR}), several restricted models can be identified, each of them describing how the present state of the whole network system depends on the past states of one or more subsystems. For the generic subset of $M$ processes $X^M \subset X$, the restricted model relating $X_n$ to $X^{M}_{<n}$ is formulated as
\begin{equation}
   X_n = \sum_{k=1}^{q}\mathbf{A}^{M}_k X^{M}_{n-k} + W^{M}_n ,
   \label{VARred}
\end{equation}
where the model order $q$ is theoretically infinite, the $N \times M$ coefficient matrix $\mathbf{A}^M_k$ relates the present of $X$ to the past of $X^M$ assessed at lag $k$, and $W^M$ is an $N$-dimensional noise process with covariance matrix $\Sigma_{W^M}$.
Then, from the reduced model (\ref{VARred}), the MI between the present of $X$ and the past of $X^M$ is computed in analogy to (\ref{PIgauss}) as
\begin{equation}
    I(X_n;X^M_{<n})=\frac{1}{2} \log \frac{|\Sigma_X|}{|\Sigma_{W^M}|}.
   \label{MIgauss}
\end{equation}
Any reduced reduced model in the form of (\ref{VARred}) describing the contribution of a given set of processes to the future evolution of the network can be identified from the parameters of the full model (\ref{VAR}) by the procedure described in detail in \cite{mijatovic2024network}; briefly, the procedure first infers the time-lagged covariance structure of the full model (\ref{VAR}) by solving a discrete-time Lyapunov formulation of its Yule-Walker equations, 
and then prunes the covariance matrices to derive the covariance structure of the restricted model (\ref{VARred}) as well as its parameters via the solution of the relevant Yule-Walker equations.

The procedure described above has the important advantage that it only relies on the identification of the parameters of the full model (\ref{VARred}), i.e. $\mathbf{A}_k$ and $\Sigma_U$, which is typically performed applying the ordinary least squares method to the available set of time series and selecting the order $p$ via information-theoretic criteria \cite{faes2012measuring}. Then, the parameters of any restricted model as in (\ref{VARred}) can be obtained, setting an arbitrarily large order $q$, without the need to re-identify such models from subsets of time series. Finally, these parameters matrices are used  to compute the many MI terms needed in (\ref{WMSdec}) to compute $\Delta_{\mathrm{WMS}}$, and in (\ref{Red}), (\ref{PIDatoms}) and (\ref{PIDdec}) to compute respectively the redundancy function, the information associated to each atom, and the coarse-grained PID terms.


\subsection{Theoretical Example}
The PrID framework can be illustrated in network Gaussian systems whose evolution over time is mapped by a VAR process in the form of (\ref{VAR}), for which the WMS and PID decompositions can be computed analytically from the theoretical values imposed for the VAR parameters.
Here, we consider a VAR model of order $p=2$ fed by $N=3$ independent Gaussian noises with zero-mean and unit variance ($\Sigma_W=\mathbf{I}_3$). The dynamics of the output processes $X=\{X^1,X^2,X^3\}$ result from the coefficients $a_i$ modulating the self-dependencies (set at lag 2) of the process $X^i$, and from the coefficients $c_{ji}$ modulating the causal interactions set at lag 1 from $X^i$ to $X^j$ ($i,j \in \{1,2,3\}$, $i \neq j$; connectivity matrices in Fig. 2a).

The simulation was performed at varying the strength of the causal coupling from $X^1$ to $X^2$ and $X^3$ modulated by the parameters $c_{21} \in [0,0.5]$ and $c_{31}  \in [0,0.5]$, with fixed internal dynamics of the three processes ($a_1=a_3=0.5$, $a_2=0.15$) and bidirectional interactions between $X^2$ and $X^3$ ($c_{32}=0.15$, $c_{23}=0.5$).
The results of the WMS and PID decompositions of the PI are reported in Fig. 2b-d. In the case of bidirectional coupling $X^2 \leftrightarrow X^3$ with isolated $X^1$ ($c_{21}=c_{31}=0$), the PID highlights how the PI of $X_n$ is composed by unique information from $X^3$ and prevalence of synergy over redundancy.
When a stronger causal coupling from $X^1$ to $X^2$ and/or to $X^3$ is simulated by increasing $c_{21}$ and/or $c_{31}$, the growing amount of the PI due to $X^1$ determines a decrease of the unique information contributed by $X^3$ and a rise of the unique information contributed by $X^1$ (Fig. 2d). Moreover, the nature of the higher-order interactions resulting from such stronger couplings depends on which arm of the system is activated originating from $X^1$: if the coupling is increased towards $X^2$ which produces less PI (as $a_2$ and $c_{32}$ are small), then the net synergy is preserved and tends to increase; if the coupling is increased towards $X^3$ which is more connected (as $a_3$ and $c_{23}$ are large), then the net synergy progressively vanishes and net redundancy is established (Fig. 2c). Remarkably, the presence of net synergy cannot be inferred using the WMS decomposition (Fig. 2a).

\begin{figure} [t!]
    \centering
    \includegraphics[scale= 1.02]{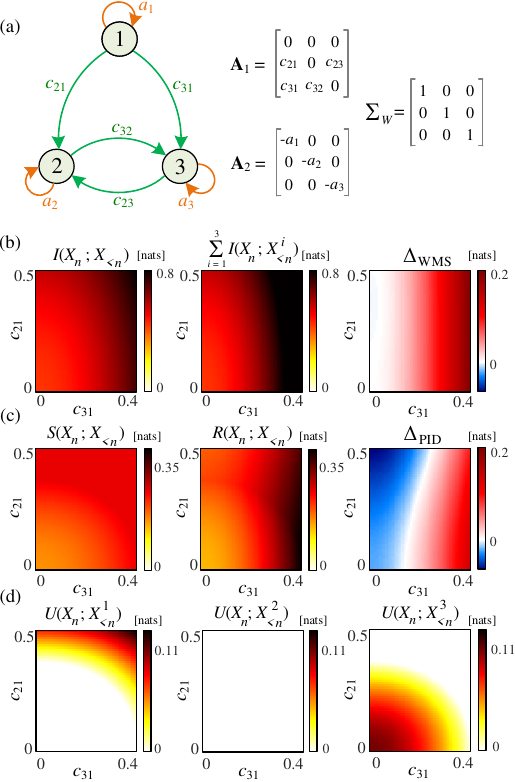}
    \caption{Predictive Information decomposition in simulated Gaussian system composed of three interacting processes $X^1, X^2, X^3$. (a) Schematic of the self-dependencies and causal interactions (orange and green arrows), with matrices of the imposed VAR model parameters. (b) Whole-minus-sum (WMS) decomposition showing the color-coded values of the PI (left), the sum of the three MIs relevant to the individual sources (middle), as well as their difference (right). (c,d) Partial Information Decomposition (PID) of the PI evidencing the color-coded values of the synergistic and redundant components and of their difference (c, from left to right) and of the three unique components (d). Each information measure in panels (b-d) is computed as a function of the two coupling parameters linking $X^1$ to $X^2$ and $X^3$.}
    \label{res_sim}
\end{figure}


These results highlight the complexity of the patterns of multivariate interactions even in a simple low-dimensional system with linear dynamics. Confirming preliminary investigations in bivariate systems \cite{faes2023investigating}, we find that the PI tends to increase with the parameters modeling self-dependencies in a process and cross-dependencies between processes. Moreover, several tests with varying network configurations reveal that redundancy arises from the duplication of information originating from the same source (e.g., via common-drive or cascade directed interactions), while synergy emerges when more sources contribute to the dynamics of the same target process (e.g. through configurations with colliders). Therefore, emergent behaviors are favored by the presence of multiple causal effects (either causal interactions or self-dependencies) pointing to the same target.
Our simulations also highlight that, although the WMS decomposition provides equal synergy/redundancy balance than the PID when $N=2$ (see (\ref{double_counting})), the PID is recommended to investigate emergent behaviors in the presence of more than two processes as it avoids double counting redundancy in the computation of the balance.

\section{Application to Physiological Networks}

The proposed PrID framework was applied to analyze the dynamics of the physiological network underlying the short-term regulation of the cardiovascular and cardiorespiratory systems, assessed in the resting state and during postural stress \cite{javorka2017basic,faes2016information}. The analyzed processes map the temporal evolution of important cardiovascular variables, i.e. heart period (HP), systolic and diastolic arterial pressure (SAP, DAP), and respiration (RESP). Thse variables are known to dynamically interact, at the time scale of heartbeats, to preserve the homeostatic function of the human organism, reflecting a variety of neuroautonomic and mechanical control mechanisms \cite{malpas2002neural,javorka2017basic}.
Here, we study how the overall behavior of cardiovascular and cardiorespiratory networks emerges from high-order interactions decomposing the predictive information through the approach defined in Sect. II and implemented in Sect. III.

\subsection{Experimental Protocol and Data Analysis}

We consider a historical dataset comprising the variability series of HP, SAP, SAP and RESP measured in 61 young healthy subjects (37 females, $17.5 \pm 2.4$ years) who underwent a protocol including a phase of relaxation in the resting supine position (SU), followed by a phase of postural stress in the $45^{\circ}$ upright position (UP) reached by passive head-up tilt. The time series were synchronously measured from the electrocardiogram (ECG, lead II on the thorax), finger arterial pressure (AP, recorded via the photoplethysmographic volume-clamp technique) and respiratory signal (recorded via inductive plethysmography) acquired simultaneously (sampling frequency of 1 KHz) as follows: the present HP sample, $H_n$, was taken as the duration of the interval between the $n^{\mathrm{th}}$ and $(n+1)^{\mathrm{th}}$ heartbeats detected as the time occurrences of the R-peaks of the ECG; the present SAP and DAP samples, $S_n$ and $D_n$, were taken as the maximum value of the AP waveform inside the interval $H_n$ and as the following minimum AP value; the present RESP sample, $R_n$, was taken as the value of the respiratory signal taken at the onset of the interval $H_n$.
After measurement, segments of consecutive 300 samples of the four series were selected synchronously for each subject in both the SU and UP conditions. To remove artifacts and favor the fullfillment of stationarity, the time series were pre-processed detecting and removing outliers, and applying a zero-phase IIR high-pass filter with cutoff frequency of 0.0107 cycles/beat (optimized to remove slow trends while not interfering with physiological dynamics). Full details about the study group, experimental protocol and time series extraction are provided in \cite{javorka2017basic,faes2016information}.

Here, the time series were grouped in triplets to study cardiovascular interactions through the variability of SAP, DAP and HP, vascular-respiratory interactions through the variability of SAP, DAP and RESP, and cardiovascular-respiratory interactions through the variability of SAP, HP and RESP. 
To this end, the PrID framework was applied three times for each subject and experimental condition, identifying the network process $X=\{X^1,X^2,X^3\}$ in the three cases as $X=\{S,D,H\}$, $X=\{S,D,R\}$, and $X=\{S,H,R\}$.
In each case, the three analyzed time series were reduced to zero mean and then fitted with a VAR model in the form of 
(\ref{VAR}); the model was identified performing least squares estimation of the parameters $\mathbf{A}_k$ and $\Sigma_U$, setting the order $p$ according to the Bayesian Information Criterion \cite{schwarz1978estimating}. The estimated parameters were used together with the estimate of the covariance $\Sigma_X$ to compute the PI via (\ref{PIgauss}), 
as well as to identify the restricted models (\ref{VARred})  needed for the computation of the MI terms in the WMS and PID decompositions according to the procedure described in Sect. \ref{sec:Computation}; the order of the restricted models was set to $q=20$, a value typically sufficient to capture the decay of the correlations for stable VAR processes \cite{faes2016information,mijatovic2024network}. 

The statistical significance of the PI and of each term in the WMS and PID decompositions was assessed using a test based on surrogate data. Specifically, for each subject, condition and analyzed triplet, surrogates of the three analyzed time series were built by shuffling randomly the samples of the series; the same random permutation was applied to the three series, in order to maintain correlations at zero lag which are not quantified by the PI. One hundred surrogate time series were generated in this way, and then a significance test based on percentiles was run with $5\%$ significance: any considered non-negative measure (e.g., the PI) was deemed as statistically significant if its estimate obtained for the original series was above the $95^{\mathrm{th}}$ percentile of the distribution derived from the surrogates; when applied to the synergy/redundancy balance $\Delta_{\mathrm{WMS}}$ and $\Delta_{\mathrm{PID}}$, the two-tail implementation of the test was performed comparing the original measure with the $2.5^{\mathrm{th}}$ and $97.5^{\mathrm{th}}$ percentiles of the surrogate distribution.

\subsection{Results and Discussion}

The results of the PrID decomposition performed using the WMS and PID approaches are reported in {Figs. \ref{res_real_fig1}-\ref{res_real_fig3}, showing the distributions across subjects of the various measures of information dynamics computed for each selected triplet of time series (i.e., $\{S,D,H\}$, $\{S,D,R\}$ and $\{S,H,R\}$) in the two analyzed conditions SU and UP. For each measure, the percentage of subjects (out of 61) for which the measure was detected as statistically significant according to surrogate data analysis is reported above the distribution. Moreover, statistically significant variations in the transition from SU to UP, assessed for each measure by the Wilcoxon signed rank test, are marked with the   $\#$ symbol.

\begin{figure} [t!]
    \centering
\includegraphics[scale= 0.87]{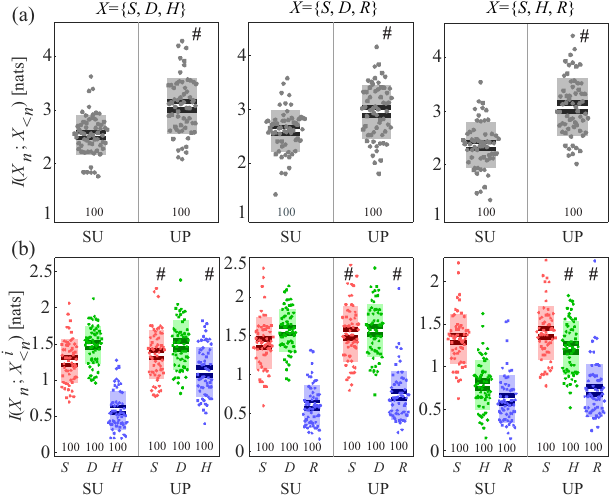}
    \caption{Whole-minus-sum decomposition of the predictive information of physiological networks.    
    Each panel reports the distributions (boxplot and individual values) of the PI between the present and the past network states (a) or the MI between the present network state and the past of the $i^{\mathrm{th}}$ unit ($i=1$: red; $i=2$: green; $i=3$: blue) (b), computed in the supine (SU) and upright (UP) conditions for the cardiovascular (left panels), vascular-respiratory (middle panels) and cardiovascular-respiratory (right panels) networks.
    The percentage of subjects for which each measure was detected as statistically significant is reported below its distribution. Statistically significant differences between SU and UP ($p<0.05$) are marked with $\#$.}
\label{res_real_fig1}
\end{figure}

\subsubsection{WMS approach to PI analysis} Fig. \ref{res_real_fig1} reports the results of the analysis based on the standard MI measures used to implement the WMS decomposition.
The PI $I(X_n;X_{<n})$ was high and statistically significant for all triplets of series in both conditions, and increased significantly moving from SU to UP (Fig. \ref{res_real_fig1}a)}.
The MI contributions to the overall dynamics computed using only one predictor time series were also statistically significant in all cases (Fig. \ref{res_real_fig1}b); the MI terms were generally higher for the AP processes $D$ and $S$ than for the heart period and respiratory processes $H$ and $R$, and showed a tendency to increase from SU to UP that was statistically significant for $S$, $H$ and $R$.

These results document the presence of significant predictive information in the analyzed cardiovascular and respiratory networks, and their strengthening induced by postural stress. The described behaviors confirm those observed in previous studies showing that cardiovascular and respiratory processes exhibit predictable dynamics induced by the action of several physiological control mechanisms, and that postural stress reduces the complexity of such processes investigated through univariate and multivariate analyses \cite{valente2018univariate}.


\begin{figure} [t!]
    \centering
\includegraphics[scale= 0.87]{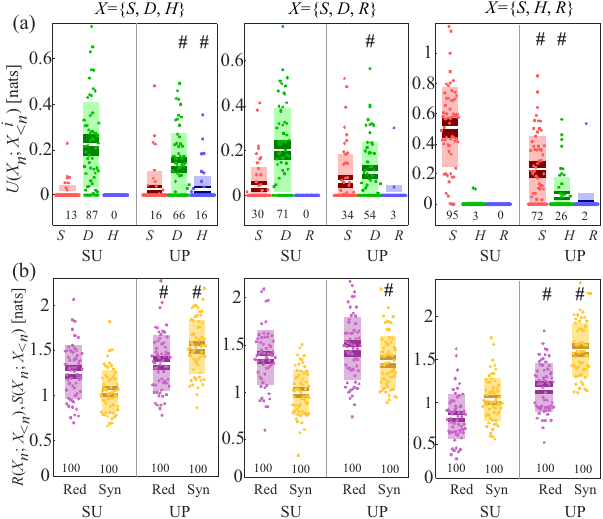}
    \caption{Partial information decomposition of the predictive information of physiological networks.    
    Each panel reports the distributions (boxplot and individual values) of the unique information provided to the present network state by the past of the $i^{\mathrm{th}}$ unit ($i=1$: red; $i=2$: green; $i=3$: blue) (a),
    or the redundant (violet) and synergistic (yellow) information provided to the present network state by the past state of the units (b), 
    computed in the supine (SU) and upright (UP) conditions for the cardiovascular (left panels) vascular-respiratory (middle panels) and cardiovascular-respiratory (right panels) networks.
    The percentage of subjects for which each measure was detected as statistically significant is reported below its distribution. Statistically significant differences between SU and UP ($p<0.05$) are marked with $\#$.}
\label{res_real_fig2}
\end{figure}

\subsubsection{PID approach to PI decomposition}
The contributions of the parts to the dynamics of the whole system describing the analyzed networks emerge more evidently from the analysis of the PID measures reported in Fig. \ref{res_real_fig2}. In particular, the unique information evidenced better than the MI the role of the individual processes in predicting the network evolution (Fig. \ref{res_real_fig2}a): in the SU condition, the largest unique contributions were given by DAP to the networks $\{S,D,H\}$ and $\{S,D,R\}$, and by SAP to the network $\{S,H,R\}$; moving from UP to SU the unique contributions became more balanced, with significant decreases for DAP in the networks $\{S,D,H\}$ and $\{S,D,R\}$, significant decrease for SAP in the network $\{S,H,R\}$, and significant increase for HP in the networks $\{S,D,H\}$ and $\{S,H,R\}$.
Physiologically, the prominent unique contribution of the DAP and SAP variables to the state of the overall system can be explained considering that AP is the primary target of cardiovascular regulation \cite{parati2018blood}, and its main oscillations due to self-regulation loops prevail in physiological variability \cite{malpas2002neural}. On the other hand, the increased importance of unique HP contributions during postural stress can be related to the fact that also HP becomes a target of the regulation in this condition, due to an enhancement of sympathetic activity and baroreflex control which emphasize the predictable oscillations of heart rate \cite{faes2013mechanisms,valente2018univariate}.

As regards the redundant and synergistic dynamics, both these contributions to the PI were statistically significant and higher than the unique terms for all triplets in both conditions; moreover, both redundancy and synergy increased significantly moving from SU to UP (Fig. \ref{res_real_fig2}b).
Thus, while unique contributions of the vascular dynamics (and also the cardiac dynamics during tilt) play a role in driving overall variability, the highest predictive capacity is provided by redundant and synergistic interactions, which are emphasized by postural stress. The nature of these high-order interactions, which have been documented in several previous studies \cite{porta2017quantifying,porta2017assessingred,javorka2018towards}, can be better elucidated looking at the synergy/redundancy balance.

\subsubsection{Synergy/redundancy balance}
\begin{figure} [t!]
    \centering
\includegraphics[scale= 0.87]{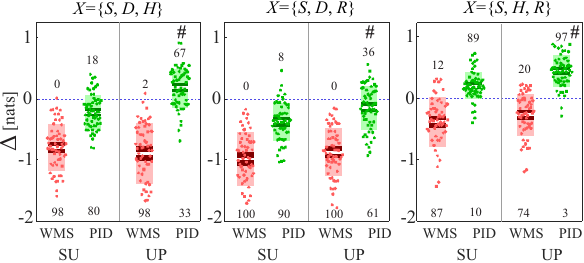}
    \caption{Synergy/redundancy balance of the predictive information of physiological networks.    
    Each panel reports the distributions (boxplot and individual values) of the difference between synergistic and redundant information provided to the present network state by the past state of the units, computed in the supine (SU) and upright (UP) conditions for the cardiovascular (left panels), vascular-respiratory (middle panels) and cardiovascular-respiratory (right panels) networks.
    The percentage of subjects for which each measure was detected as statistically significant is reported above (below) its distribution in the case of significant net synergy (redundancy). Statistically significant differences between SU and UP ($p<0.05$) are marked with $\#$.}
\label{res_real_fig3}
\end{figure}

Fig. \ref{res_real_fig3} reports the distribution of the synergy/redundancy balance computed using the WMS and PID, showing how the two approaches evidence markedly different results. Using the WMS decomposition, the balance measure is largely negative for all triplets, suggesting the dominance of redundancy, and is not modulated by the transition from SU to UP. On the other hand, the PID approach revealed different patterns for the three analyzed physiological networks: for $\{S,D,H\}$, redundancy was prevalent during SU and synergy was prevalent during UP; for $\{S,D,R\}$, redundancy was prevalent in both conditions; for $\{S,H,R\}$, synergy was prevalent in both conditions. Moreover, the synergy/redundancy balance increased significantly moving from SU to UP in all the analyzed networks.

The patterns of net redundancy observed using the WMS measure are in agreement with several previous findings consistently documenting a large prevalence of redundancy over synergy across different methods of assessment,  cardiovascular variables analyzed and experimental conditions \cite{porta2017quantifying,porta2017assessingred,javorka2018towards,porta2018paced,faes2021information,porta2021categorizing}. These findings were commonly explained by the evidence of common-drive modulatory effects, e.g. due to respiration and/or sympathetic activity, impacting the short-term cardiovascular regulation.
However, our results suggest that the balance between synergy and redundancy, if computed using the PID and considering the collective present state and the individual past states as target and source variables, highlights previously unreported, network-specific modes of cardiovascular interaction which can be related to causal emergence.
In vascular-pulmonary networks probed by SAP, DAP and RESP, the balance is shifted towards dominant redundancy, likely reflecting the unidirectional nature of the mechanical influences of respiration on arterial pressure (via the effects on lung volume and intrathoracic pressure \cite{saul1991transfer}) and of diastolic over systolic pressure (via the effect of end-diastolic left ventricular volume on cardiac contraction \cite{javorka2017basic}). 
In cardiovascular-respiratory networks probed by SAP, HP and RESP, the balance is shifted towards dominant synergy, pointing to the presence of emergent collective behavior; this behavior can be explained by the coexistence of several mechanisms impacting spontaneous variability, which include again the mechanical influence of RESP on SAP \cite{saul1991transfer}, but also respiratory effects on HP resulting in the well-known respiratory sinus arrhythmia \cite{hirsch1981respiratory} as well as reciprocal influences between HP and SAP manifested through baroreflex feedback (direction SAP$\rightarrow$HP) and mechanical feedforward (direction HP$\rightarrow$SAP) mechanisms\cite{malpas2002neural,faes2013mechanisms}.
Finally, in cardiovascular networks probed by SAP, DAP and HP, the balance moves from predominant redundancy at rest to predominant synergy during tilt, reflecting the enhancement of closed-loop interactions between SAP and HP and the decrease of respiration-related effects and the baroreflex activation occurring with postural stress \cite{faes2013mechanisms,javorka2017basic}. These two physiological effects are likely the factors underlying the increase of synergy observed for all network configurations, which is regarded as an indication of stronger emergence capacity induced by the orthostatic stress.



\section{Conclusions}

This work introduces an information-theoretic tool to quantify emergent high-order behaviors in dynamic network systems mapped by multivariate physiological time series. Taking inspiration from the recently introduced theory of causal emergence \cite{rosas2020reconciling}, our approach is based on computing the predictive information of the observed network dynamics and on dissecting it through partial information decomposition. The main features of this approach are the focus on temporal evolution, the investigation of collective behaviors, and the explicit account of high-order interactions through distinct synergy and redundancy measures. These properties make our tool intrinsically different from most of the existing approaches proposed to decompose multivariate information in network physiology: standard WMS or PID metrics \cite{timme2014synergy} applied to random variables do not account for temporal information; joint transfer entropy approaches \cite{porta2017quantifying}, even when based on PID \cite{faes2021information}, do not focus on collective dynamics; integrated information measures \cite{barrett2011practical,oizumi2016unified} do not elicit high-order interactions explicitly; dynamic high-order interaction measures like the O-information rate \cite{faes2022new} cannot separate synergy from redundancy.

The proposed approach allows quantifying both the individual contribution of each network unit to the overall dynamics via unique information, and the collective high-order contributions via redundant and synergistic information. Following the principles of causal emergence \cite{rosas2020reconciling}, emergent behaviors are assessed by synergy, and quantified in a strong form when synergy prevails over redundancy. Our simulations demonstrate that the presence of a positive synergy/redundancy balance is a proxy of rich integrated dynamics determined by coexisting multiple causality patterns and internal dynamics, while conditions of unidirectional and common drive coupling are reflected by net redundancy. Crucially, we show that the separation between synergy and redundancy guaranteed by PID is necessary to avoid underestimation of their balance in systems with more than two units.

When applied to physiological networks mapping the beat-to beat cardiovascular and respiratory variability,  the proposed method puts in evidence modes of interactions never observed before, which may help in both disclosing the underlying physiological control mechanisms and identifying biomarkers with potential clinical relevance. Our main result is that the synergy/redundancy balance assessed in cardiovascular and respiratory networks is a hallmark of integrated short-term control in these networks, and rises in response to an activation of the sympathetic nervous system. This established link with autonomic regulation suggests that the proposed measures
can be fruitfully exploited to classify groups of patients featuring different degrees of autonomic impairment. Moreover, given the close connection of the proposed framework with the concept of causal emergence \cite{rosas2020reconciling}, we advocate further studies investigating features that can exhibit emergent behavior with respect to cardiovascular variability.

\section{Acknowledgments}
The Authors thank Michal Javorka for sharing the dataset of cardiovascular variability series used for the analyses reported in this work.

This research was supported by the project "HONEST - High-Order Dynamical Networks in Computational Neuroscience and Physiology: an Information-Theoretic Framework”, Italian Ministry of University and Research (MUR, PRIN 2022, code 2022YMHNPY, CUP: B53D23003020006)


\bibliographystyle{IEEEtran}
\bibliography{references}
\end{document}